\documentclass{elsart}
\usepackage{natbib}
\usepackage{graphicx}

\usepackage{amssymb}

\begin{document}

\begin{frontmatter}



\title{Contact switching as a control strategy for epidemic outbreaks}


\author{Sebasti\'an Risau-Gusman}

\address{Consejo Nacional de Investigaciones Cient\'{\i}ficas y T\'ecnicas,
Centro At\'omico Bariloche, 8400 San Carlos de Bariloche, R\'{\i}o Negro, Argentina}

\author{Dami\'an H. Zanette}

\address{Consejo Nacional de Investigaciones Cient\'{\i}ficas y T\'ecnicas,
Centro At\'omico Bariloche and Instituto Balseiro,
8400 San Carlos de Bariloche, R\'{\i}o Negro, Argentina}

\begin{abstract}
We study the effects of switching social contacts as a strategy to
control epidemic outbreaks. Connections between susceptible and
infective individuals can be broken by either individual, and then
reconnected to a randomly chosen member of the population. It is
assumed that the reconnecting individual has no previous information
on the epidemiological condition of the new contact. We show that
reconnection can completely suppress the disease, both by continuous
and discontinuous transitions between the endemic and the
infection-free states. For diseases with an asymptomatic phase, we
analyze the conditions for the suppression of the disease, and show
that --even when these conditions are not met-- the increase of the
endemic infection level is usually rather small. We conclude that,
within some simple epidemiological models, contact switching is a
quite robust and effective control strategy. This  suggests that it
may also be an efficient method in more complex situations.
\end{abstract}

\end{frontmatter}

\section{Introduction}
\label{Intro}

Mathematical models have been applied to the study of infectious
diseases since more than a century ago. The last four decades have
witnessed a burst of interest in quantitatively understanding the
transmission dynamics of a large number of diseases \citep{AM}. One
of the key aims of epidemiological mathematical models, and
certainly the most relevant in terms of policy making, is the
assessment of the effectiveness of control strategies to curb
disease spreading. For many infectious diseases, the most widespread
prevention measure is mass vaccination. However, if for a given disease
vaccines are not known, or vaccination is not effective,
other control measures have to be adopted. These mostly consist in
some form of isolation of the infective individuals. Isolation from
an individual's social environment, however, can have serious
psychological effects such as depression and stress \citep{HGRPGS}.
An even more drastic measure is quarantine, where individuals {\em
assumed} to have been in contact with infective people are
temporarily isolated from the rest of society. Daniel Defoe gives a
forceful description of this practice and its tragic effects during the
great plague of London (1665-1666) in his {\it Journal of the Year
of the Plague}. A different strategy, that has helped preventing the
spread of some sexually transmitted diseases, is contact tracing,
where all the partners of an infective individual are located to be
informed about the possibility of infection and, eventually, given
adequate treatment. It has been argued, however, that this practice
violates the right to privacy. In the case of HIV, for instance, it
has sometimes met with considerable opposition \citep{BT}. It must
also be mentioned that all these measures require widespread
governmental action and, frequently, allocation of substantial
resources \citep{AB}.

Recently, it has been suggested that a different kind of strategy,
implemented by the individuals themselves, could give surprisingly
positive results in controlling disease spreading \citep{GDdLB,ZR-G}.
The basic idea is that contact between non-infective and infective
acquaintances must be systematically avoided. This can be
implemented by either non-infective or infective individuals, or by
both. The effects of this simple strategy have been explored by
means of network epidemiological models. In these models, each
member of a population --hereafter, an {\it agent}-- occupies a node
of a network \citep{KE}. Neighboring agents, which are connected by
network links, can interact, and their epidemiological states
(susceptible, infective, recovered, etc.) thus change. For instance,
the interaction of a susceptible agent and an infective agent may
lead, with a certain probability, to the infection of the former.
The set of neighbors of a given agent represents the individual's
acquaintances.

In the framework of network models, the mechanism of avoiding
contact with infective acquaintances amounts to breaking the links
between susceptible and infective agents. If the social connectivity
is nevertheless to be preserved, those broken links must be replaced
by new connections. In the SIS (susceptible $\to$ infective $\to$
susceptible) epidemiological model analyzed by \citet{GDdLB}, it is
susceptible agents who break contacts with their infective
neighbors, what necessarily leads to the isolation of infective
agents, and new connections are established only with other
susceptible agents. Implicitly, this assumes that the
epidemiological state of all agents is known to any other agent,
irrespectively of whether they are connected or not. More
realistically, \citet{ZR-G} consider links that are broken and reconnected to
randomly chosen members of the population. As described in the next
section, we adopt here the same viewpoint. Furthermore, in order to
avoid straightforward isolation of infective agents, in Section
\ref{infectives} we generalize the model in order to also allow
infective individuals to switch connections from their susceptible
neighbors to new contacts chosen at random. We show that, even under
these less stringent assumptions on the mechanism of link switching,
the outcome is still an effective reduction in the probability of
infection, which can eventually lead to the complete suppression of
the disease. In Section \ref{asymptomatics}, we study the case where
the disease has an asymptomatic period, during which infective
individuals are not perceived as such but appear as being
susceptible, even to themselves \citep{FRAF}. In this situation, it
may happen that an asymptomatic agent switches contact from an
infective neighbor to a susceptible agent, thereby increasing the
probability of transmission. We find that, if the asymptomatic
period is not very long, large enough switching rates can still
suppress the infection. Results are summarized and discussed in the
last section.

\section{SIS network model with switching links}
\label{basic}

We consider a population formed by $N$ agents, situated at the nodes
of a network. At a given time, each of them can be either
susceptible ($S$) or infective ($I$). Initially, each agent has
exactly $k$ neighbors and, thus, the network has $kN/2$ links. The
initial contacts are chosen at random, so that the resulting network
is a random regular graph. The corresponding degree distribution,
i.e. the distribution of the number of neighbors per agent, is thus
a delta-like function centered at $k$ neighbors.

The infection is transmitted from infective agents  to their
susceptible neighbors, who become immediately infective ($S\to I$),
at rate $\lambda$. Moreover, infective agents get cured at rate
$\gamma$. Upon cure, they return to the susceptible state ($I \to
S$) and, subsequently, can be infected again.  The diseases
described by the SIS model are those that do not confer immunity
against subsequent infections, and which have very short latent
periods. Models of this class have proven particularly useful for
the study of chlamydia and gonorrhea \citep{HY,KvDS,GSG,TAGGME}.

Contact switching, in turn, is modeled as follows. All links joining
susceptible agents (or, for short, {\em susceptibles}) with
infective agents (or {\em infectives}) are broken at rate $r$, and
the corresponding susceptible agent is then connected to a randomly
chosen agent in the population. Self-connections and multiple
connections between pairs of agents are avoided.

Taking into account the above dynamical rules, and using the pair
approximation discussed in the Appendix, it is possible to write
down evolution equations for the variables  $n_A (t)$ and $m_{AB}
(t)$,  with $\{ A,B \} \equiv \{ S,I \}$. They represent,
respectively, the fraction of agents in state $A$ and the fraction
of links joining agents in states $A$ and $B$. The equations read
\begin{eqnarray}
\dot n_I  &=& -n_I \gamma + \lambda m_{SI}, \nonumber \\
\dot m_{SI}  &=& 2\gamma m_{II} - \gamma m_{SI} + \lambda
kK  m_{SI} \frac{2m_{SS}-m_{SI}}{n_S}-\lambda m_{SI}-r n_S m_{SI} ,\label{ecs1} \\
\dot m_{II} &=& -2\gamma m_{II}+ \lambda m_{SI} + 2 \lambda k K
m_{SI} \frac{m_{SI}}{n_S}. \nonumber
\end{eqnarray}
As the total number of agents and links is preserved, this system is
completed by the relations $m_{II}+m_{SI}+m_{SS}=1$ and $n_I+n_S=1$.
The constant $K=(k-1)/k$ derives from the pair approximation. The high-connectivity limit, $K=1$ has been considered by \citet{GDdLB}.

In the remaining of this article we concentrate on the equilibrium
properties of our system, obtained by equating the right-hand side
of Eqs. (\ref{ecs1}) to zero. After some algebra, we find that the
equilibrium fraction of infectives $n_I$ satisfies the fourth-degree
polynomial equation
\begin{eqnarray}
n_I [-{\tilde r}(1-n_I)^3 &-& (1-n_I)^2 -K^2 n_I^2 +  \nonumber \\
&+& K(1-n_I)(k {\tilde \lambda} - (2+ {\tilde \lambda} )n_I)]=0.
\label{basic_pol}
\end{eqnarray}
The parameters ${\tilde r}= r/ \gamma$ and ${\tilde
\lambda}=\lambda/ \gamma$ result from rescaling time in Eqs.
(\ref{ecs1}), so that the time unit is the recovery time
$\gamma^{-1}$. The infection-free state, $n_I=0$, is a trivial
solution of Eq. (\ref{basic_pol}) for all values of the parameters.
As expected, in fact, the infection cannot spontaneously appear in
the absence of infectives. A second solution of the equation is
always larger than unity. It therefore has no biological meaning,
and will not be further discussed. As for the remaining solutions,
it can be shown that for every value of the reconnection parameter
${\tilde r}$, there is a critical value of infectiveness ${\tilde
\lambda}_c$ at which a transition from the infection-free (${\tilde
\lambda}<{\tilde \lambda}_c$) to the endemic state (${\tilde
\lambda}>{\tilde \lambda}_c$), with persisting infection, occurs.

To determine the nature of this transition we perform a stability
analysis of the solutions of Eq. (\ref{basic_pol}).  When the
reconnection parameter is below the threshold ${\tilde
r_c=[(2K-1)k+1]/(2k-1)}$, the situation is similar to the case
without reconnection ($\tilde r=0$). For small ${\tilde \lambda}$
the only stable non-negative solution is $n_I=0$. At ${\tilde
\lambda}_c$, a transcritical bifurcation takes place. Namely,
another real solution becomes simultaneously positive and stable,
whereas $n_I=0$ becomes unstable. For larger values of the
infectiveness, the equilibrium fraction of infectives grows
continuously from zero.

When, on the other hand, ${\tilde r}>{\tilde r_c}$, the only real
solution for ${\tilde \lambda}<{\tilde \lambda_c}$ is $n_I=0$. Now,
the transition at the critical infectiveness ${\tilde \lambda_c}$ is
a tangent bifurcation, and two real and positive solutions appear
simultaneously. For ${\tilde \lambda}>{\tilde \lambda_c}$, the
larger of these two solutions is stable and the other one is
unstable, while the stability of $n_I=0$ does not change. We have
therefore a discontinuous transition from the infection-free state
to the endemic state. Further increasing ${\tilde \lambda}$, a
transcritical bifurcation point is reached, where the unstable
solution vanishes and exchanges stability with $n_I=0$. From then
on, there is a single non-negative stable solution. Between the two
bifurcations, the system has two stable equilibrium states, one
endemic and the other infection-free. In this bistable region, the
long-time asymptotic state is selected by the initial state of the
population. An initial fraction of infectives larger than the
unstable solution leads to the endemic state, whereas a smaller
fraction leads to the extinction of the infection. The unstable
solution thus plays the role of a threshold for the epidemics to
spread and, consequently, is sometimes called {\em breakpoint
density}  \citep{AM}.

\begin{figure} %[!h]
\centerline{\includegraphics[width=\columnwidth,clip=true]{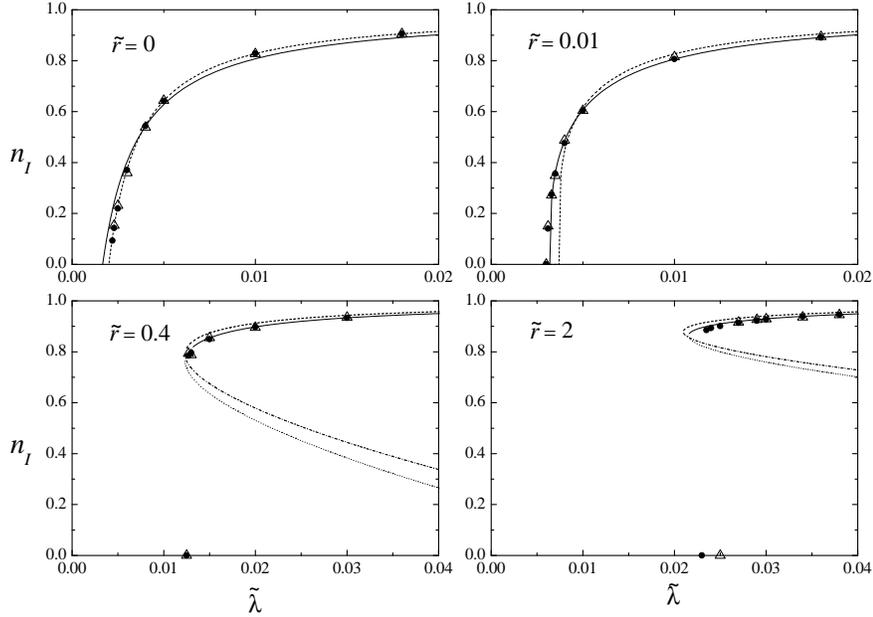}}
\caption{Equilibrium fraction of infective agents as a function of
infectivity, for $k=3$ and four different values of the reconnection
parameter $\tilde r$. Full and dashed lines are the analytical
values of $n_I$ with $K=(k-1)/k$ and $K=1$, respectively. In the
lower panels, the upper and lower branches of each curve
respectively correspond to the stable and unstable equilibrium solutions. The
lower branch represents the breakpoint density (see main text). Dots
are results of single realizations in numerical simulations for a
population of $500$ agents (triangles) and $5000$ agents (circles),
with $\gamma=1$.} \label{figure1}
\end{figure}

In Figure \ref{figure1} analytical predictions of the equilibrium
fraction of infectives, calculated from Eq. (\ref{basic_pol}), are
compared with numerical simulations of the epidemic process. As
explained below, it makes sense to consider the solutions of Eq.
(\ref{basic_pol}) fixing the parameter $K$ at both $K=(k-1)/k$ (see
the Appendix), and $K=1$. For a reconnection parameter ${\tilde
r}=0$, the analytical curve corresponding to $K=(k-1)/k$ accurately
fits the numerical results \citep{LD}. When ${\tilde r}\ne 0$, on the
other hand, for all but the smallest values of the reconnection
parameter the curves that best fit the data are those with $K=1$. A
likely reason for this discrepancy lies in the fact that, depending
on ${\tilde r}$, contact switching events generate networks with
different degree distributions. It has previously been shown that,
not unexpectedly, rewiring can lead to degree distributions that get
broader as the reconnection parameter is increased \citep{GDdLB,E}.
We have verified that the same happens in our model. It has also
been pointed out that the pair approximation is not good for
describing contact processes on networks with long-tailed degree
distributions \citep{PF}. The value $K=(k-1)/k$, in fact, results for
the pair approximation applied to regular networks with $k$
neighbors per node. A possible reason for the better fit with $K=1$
could be that, with the progressive broadening of the degree
distribution as rewiring proceeds, agents with many connections
($k\gg 1$) appear and become dominant in the epidemiological
process. These agents, usually called {\em superspreaders},  are
known to be the main vector of propagation of many diseases
\citep{L-SSKG}. Taking $K=1$ --which, as mentioned above, corresponds
to the limit of $k\gg 1$ for a regular network-- is therefore a kind
of effective approximation accounting for the role of such agents.
In the variants of the model considered below, again, better
agreement between analytical and numerical results is found for
$K=1$. Consequently, all the analytical results presented in the
following are obtained for this value of $K$.

\begin{figure} %[!h]
\centerline{\includegraphics[width=\columnwidth,clip=true]{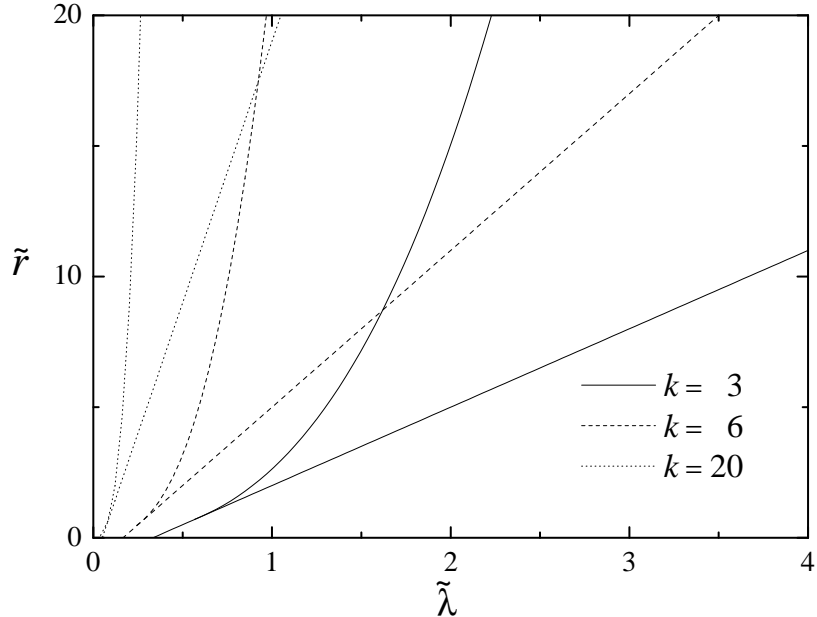}}
\caption{Phase diagram of the basic model, for three different
values of $k$, and $K=1$. For each value of $k$, the curve and the
straight line represent, respectively, the tangent and the
transcritical bifurcation. The V-shaped region between them is the
bistability zone. The left-most region corresponds to the
infection-free state, whereas the right-most region corresponds to
the endemic state.} \label{figure2}
\end{figure}

The possible asymptotic behaviors of our model can be summarized in
a phase diagram on the parameter plane $(\tilde \lambda, \tilde r)$,
as shown in Fig. \ref{figure2} for three values of the average
connectivity $k$, fixing $K=1$. For each value of $k$, the plane is
divided into three regions. In the leftmost region, corresponding to
small infectivity, the fraction of infectives asymptotically
vanishes and the population reaches the infection-free state. The
central V-shaped region  is the bistability zone, where the endemic
state is stable but the infection-free state is still reached from
initial conditions with sufficiently low levels of infectives. Note
that, as discussed above, bistability does not occur for small
reconnection parameters ($\tilde r < \tilde r_c$). Finally, for
large infectivities, we have the region where the endemic state is
the only stable equilibrium of the epidemiological process. To the
left, this region is bounded by the transcritical bifurcation line,
$\tilde \lambda_{T} = (\tilde r +1)/k$.

It is interesting to notice that  results similar to those derived
from Eq. (\ref{basic_pol}) can be obtained from the usual approach
in mathematical epidemiology, where it is assumed that an
epidemic outburst occurs as soon as an infective agent can generate
more than one secondary infection (i.e., an infection event between
two agents which involves a third, intermediary agent). This
is quantified with the basic reproductive number $R_0$,
which is the average number of secondary infections caused by a single
infective agent in an otherwise healthy population \citep{AM}. The
epidemic threshold is calculated by setting $R_0=1$. In our case, a
straightforward calculation of the basic reproductive number yields
\begin{equation}
R_0 = \frac{k{\tilde \lambda}}{{\tilde r}+{\tilde \lambda}+1},
\end{equation}
from which the epidemic threshold turns out to be ${\tilde
\lambda}_{R_0}=({\tilde r}+1)/(k-1)$. This is exactly the value at
which the transcritical bifurcation occurs in our model when
$K=(k-1)/k$. The slightly smaller result we obtain for $K=1$,
$\tilde \lambda_{T} = (\tilde r +1)/k$, is a consequence of the
above discussed correction: it is well-known that wider degree
distributions lead to smaller epidemic thresholds \citep{AM,P-SV}.

With the aim of making this basic model more realistic,  two
variants are considered in the following. In the first one,
infective agents are allowed to reconnect their contacts with
susceptibles. In the second, an asymptomatic stage in the disease is
taken into account.

\section{Reconnection of infective agents}
\label{infectives}

Our basic model can be generalized to the more realistic situation
where not only susceptibles are allowed to change their contacts,
but also infectives can break their links with susceptible
neighbours and redirect them to other members of the population.
This possibility would stand for an altruistic attitude of
infectives, in an attempt to inhibit the infection spread. To
implement this variation, we introduce a new parameter $p$. When a
link between a susceptible and an infective agent is cut, with
probability $p$ it is the infective agent who keeps the connection,
whose other end is redirected to a randomly chosen member of the
population, avoiding self- and multiple connections. With the
complementary probability, $1-p$, the connection is kept by the
susceptible agent. Thus, this variation reduces to the basic model
for $p=0$.

The evolution equations for the densities and the number of links
--fixing, as advanced above, $K=1$ in the pair approximation-- are
now
\begin{eqnarray}
\dot n_I  &=& -n_I \gamma + \lambda m_{SI}, \nonumber \\
\dot m_{SI}  &=& \gamma (2 m_{II} - m_{SI}) +\nonumber \\
&  &+ \lambda k m_{SI} \frac{2m_{SS}-m_{SI}}{n_S}
-\lambda m_{SI}-r m_{SI}[(1-p)n_S +p n_I], \label{ecs2} \\
\dot m_{II} &=& -2\gamma m_{II}+ \lambda m_{SI} + 2 \lambda k m_{SI}
\frac{m_{SI}}{n_S}, \nonumber
\end{eqnarray}
and the equation for the equilibrium fraction of infectives reads
\begin{eqnarray}
n_I \{-{\tilde r}[(1-n_I)^3&+&p(1-n_I)(2n_I-1)]\nonumber \\ &+&
{\tilde \lambda}(1-n_I)(k-n_I)-1 \}=0.
\label{polconp}
\end{eqnarray}

As in the case of $p=0$, this system undergoes a transcritical
bifurcation where the infection-free state, with $n_I=0$, becomes unstable. The
threshold infectivity at this bifurcation is $\tilde
\lambda_T=[{\tilde r}(1-p)+1]/k$. The critical value of the
reconnection parameter, above which a tangent bifurcation gives
origin to two positive solutions, is ${\tilde
r}_c=(k+1)/(1-p)(2k-1)$. Note that, for ${\tilde r}<{\tilde r}_c$,
where the epidemic threshold is given by the transcritical
bifurcation, $\lambda_T$ decreases with $p$. The reconnection of
infectives, as expected, favors infection spreading. The disease is
better controlled if only susceptibles are allowed to protect
themselves against contagion. In   the extreme case where only
infectives reconnect, the infection threshold does not depend on
${\tilde r}$.

\begin{figure} %[!h]
\centerline{\includegraphics[width=12cm,clip=true]{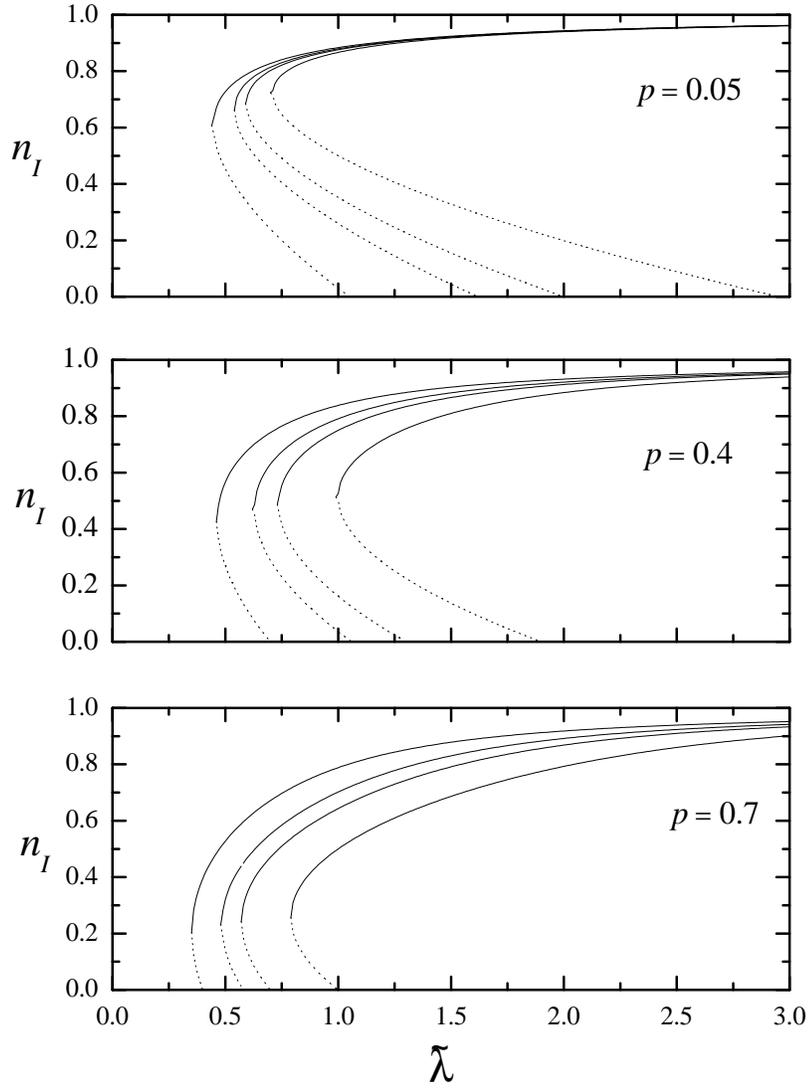}}
\caption{Bifurcation diagrams for the fraction of infectives, as a
function of the infectivity, for $k=10$ and three different values
of $p$, the probability that, for each susceptible-infective link,
it is the infective who reconnects. From left to right the curves in
each panel correspond to ${\tilde r}=10$, $16$ , $20$ and $30$. Full
and dashed lines represent stable and unstable equilibrium solutions,
respectively.} \label{figure3}
\end{figure}

\begin{figure} %[!h]
\centerline{\includegraphics[width=15cm,clip=true]{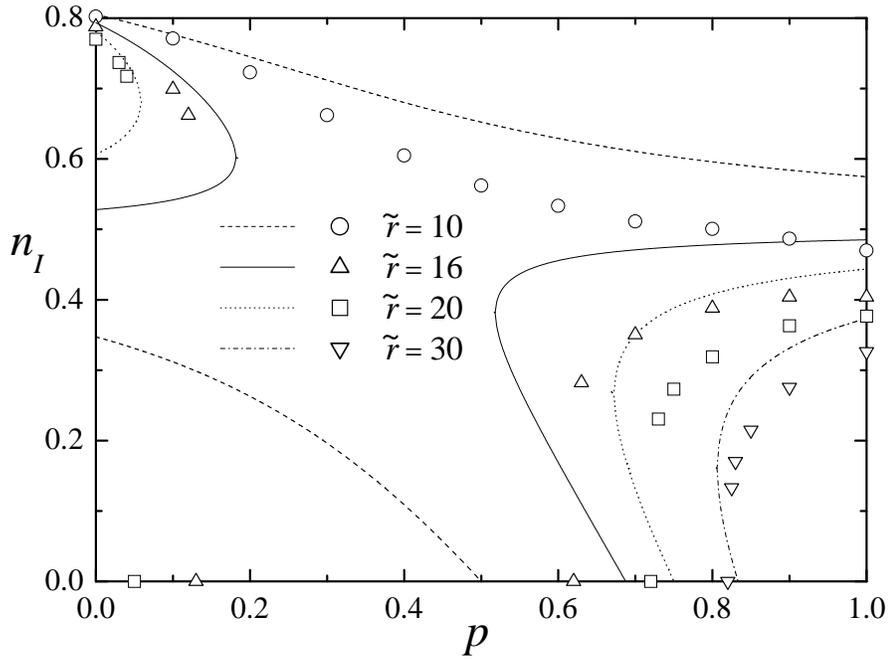}}
\caption{Bifurcation diagram for the fraction of infectives, as a
function of $p$, for $k=10$, $\tilde \lambda=0.6$ and several values of the
reconnection parameter ${\tilde r}$. The upper and lower branches of
each curve respectively correspond to the stable and unstable
solutions. Dots are simulation results for a system of $5000$
agents, with $\gamma=1$.} \label{figure4}
\end{figure}

For ${\tilde r}>{\tilde r}_c$, the behavior of the epidemic
threshold is more complicated. Figure \ref{figure3} shows that the
position of the tangent bifurcation is not monotonic with $p$. This
leads to the unusual situation depicted in Fig. \ref{figure4}. In
certain regions of parameter space, increasing $p$ first makes the
infection disappear, but a further increase of $p$ leads to the
reappearance of the endemic state through a second tangent bifurcation.
For parameter values for which the infection is not suppressed, on
the other hand, the fraction of infectives decreases monotonically.

To check that these effects are not an artifact of the mathematical
model, we have performed simulations of the corresponding contact
process \citep{LD}, whose results are represented by dots in  Fig.
\ref{figure4}. We verify that simulations display the same
qualitative behavior as the model, even though the analytical
prediction is not as good as for $p=0$ (cf. Fig. \ref{figure1}).
This is probably due to the fact that the degree distributions of
susceptibles and infectives tend to be rather broad, as discussed
above, and their respective parameters differ considerably. This
difference grows with $p$, as infectives become, on the average,
more connected than susceptibles.

\begin{figure} %[!h]
\centerline{\includegraphics[width=15cm,clip=true]{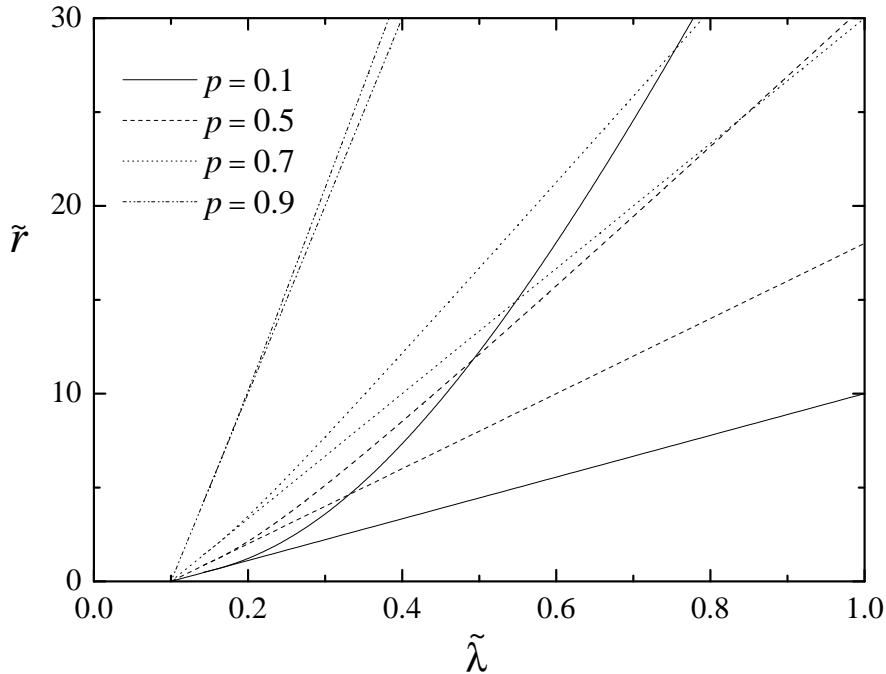}}
\caption{Phase diagram of the model with infectives who can
reconnect, for different values of $p$. For each value of $p$, the
curve and the straight line represent, respectively, the tangent and
transcritical bifurcation. Compare with Fig. \ref{figure2}.}
\label{figure5}
\end{figure}

The phase diagram in the plane $(\tilde \lambda,\tilde r)$,  shown
in Fig. \ref{figure5} for different values of $p$, reveals the
non-monotonic dependence on this parameter in the mutual crossings
of the left boundary of the bistability regions. Note that, in any
case, large enough reconnection rates always lead to the
infection-free state. In fact, reconnection  never increases the
number of susceptible-infective links. In the next section, on the
other hand, we consider a variation of the basic model where
susceptible-infective links can be created by contact switching.

\section{Effect of asymptomatic agents}
\label{asymptomatics}

Several common diseases possess an asymptomatic stage, intermediary
between the susceptible and the infective phases, during which
individuals carrying the infection do not exhibit symptoms, and are
therefore seen as non-infective by themselves and by the rest of the
population. In the asymptomatic stage, however, an infective person
may already transmit the disease by contagion. It is well known that
during the several plagues that ravaged Europe, infective people who
did not yet have visible symptoms became one of the main vectors for
the propagation of the disease, when they fled the cities. It is
therefore not unreasonable to fear that, for a disease with an
asymptomatic phase, the strategy of contact switching might
reinforce the spread of the disease, instead of curbing it.

To address these issues, in this section we analyze a model with
three types of agents, namely, $S$, $A$ (asymptomatic), and $I$. It
is basically an extension of the model studied in Section
\ref{basic}, because the interactions between susceptibles and
infectives are the same. We assume that asymptomatic agents do not
know that they are infective, and therefore act as susceptible
individuals: they reconnect their links with infective neighbors at
the same rate $r$ as susceptibles. On the other hand, asymptomatic
agents can infect their susceptible neighbors at the same rate
$\lambda$ as infectives.  Note that a main difference with the two
models analyzed above is that, now, susceptible-asymptomatic links
--which allow for the transmission of the infection-- can be created
by reconnection from asymptomatic-infective links, which join two
agents carrying the infection.

To characterize the transitions $A \to I$ and $I\to S$ we introduce
a new parameter $\alpha$. The rate at which asymptomatic agents
 become (symptomatic) infectives is $\gamma/ \alpha$,
while the rate at which infectives become susceptible is $\gamma/
(1- \alpha)$. With these definitions, the asymptomatic and infective
phases last, on the average, $\alpha/ \gamma$ and $(1-
\alpha)/\gamma$ time units, respectively. The mean total
infective period is, as before, $\gamma^{-1}$.  In the limit $\alpha
\to 0$, we recover the model of Section \ref{basic}, whereas for
$\alpha \rightarrow 1$, the model is equivalent to the usual SIS
model without reconnection.

The evolution equations for this model are
\begin{eqnarray}
\dot n_I  &=& - \frac{\gamma}{1- \alpha} n_I + \frac{\gamma}{\alpha} n_A, \nonumber \\
\dot n_A &=& -\frac{\gamma}{\alpha} n_A + \lambda k (m_{SI}+m_{SA})/2, \nonumber \\
\dot m_{SI}  &=& \frac{\gamma}{1- \alpha} (2 m_{II} - m_{SI}) +
\frac{\gamma}{\alpha} m_{SA} - \lambda
 \frac{k}{2} m_{SI} \frac{m_{SI}+m_{SA}}{n_S}-\lambda m_{SI} \nonumber \\ & & -r m_{SI} (n_S + n_A) , \nonumber \\
\dot m_{SS} &=& \frac{\gamma}{1- \alpha} m_{SI} -\lambda m_{SS}k \frac{m_{SI}+m_{SA}}{n_S} +r n_S m_{SI}, \nonumber \\
\dot m_{II} &=& -2 \frac{\gamma}{1- \alpha} m_{II}+ \frac{\gamma}{\alpha} m_{AI} ,\label{ecs3} \\
\dot m_{AI} &=& \frac{\gamma}{\alpha} (2 m_{AA}-m_{AI}) -\frac{\gamma}{1- \alpha} m_{AI} + \lambda
 \frac{k}{2} m_{SI} \frac{m_{SI}+m_{SA}}{n_S} +\lambda m_{SI} \nonumber \\ & & - r m_{AI} (n_S+n_E), \nonumber \\
\dot m_{AA} &=& -2 \frac{\gamma}{\alpha} m_{AA} + \lambda
 \frac{k}{2} m_{SA} \frac{m_{SI}+m_{SA}}{n_S} + \lambda m_{SA} +r n_A m_{AI} . \nonumber
\end{eqnarray}
The system is completed by the equations $n_S+n_A+n_I=1$ and
$m_{SS}+m_{SA}+m_{SI}+m_{AA}+m_{AI}+m_{II}=1$. In spite of the
rather imposing look of this mathematical problem, it is still
possible to find a single (sixth-order) polynomial equation for the
fraction of infectives $n_I$ at equilibrium. The equilibrium
fractions of infective and asymptomatic agents are related by the
simple equation $(1-\alpha) n_A=\alpha n_I $. Thus, at equilibrium,
$\alpha$ exactly coincides with the fraction of agents carrying the
infection which do not exhibit symptoms. This, in turn, makes it
possible to find the equilibrium fraction of susceptibles, $n_S =
1-n_I/\alpha$.

As in the models studied in Sections \ref{basic} and
\ref{infectives}, the infection-free state, $n_I=0$, is an
equilibrium solution for any value of the parameters. The
transcritical bifurcation, at which $n_I=0$ changes its stability,
occurs now at a critical infectivity ${\tilde \lambda}_T = [1+(1-
\alpha) {\tilde r}]/k[1+(1- \alpha) \alpha {\tilde r}]$. As before,
${\tilde \lambda_T}$ grows monotonically with ${\tilde r}$. Here,
however, this growth is not unbounded, and the critical infectivity
tends to ${\tilde \lambda}_{\infty} = (k \alpha)^{-1}$ for ${\tilde
r} \to \infty$. In other words, for ${\tilde \lambda}> {\tilde
\lambda}_{\infty}$ no reconnection rate is large enough to suppress
the disease. This can be understood by inspecting the basic
reproductive number corresponding to the present epidemiological
process,
\begin{equation}
R_0 = \frac{k\tilde \lambda}{\tilde \lambda+\alpha^{-1}}+
\frac{k\tilde \lambda}{\tilde r+\tilde \lambda+(1-\alpha)^{-1}} .
\end{equation}
The two terms in the right-hand side of this equation arise from the
number of secondary infections respectively generated during the
asymptomatic and symptomatic phases. As expected, only the latter
depends on the reconnection parameter. But, for ${\tilde \lambda}>
{\tilde \lambda}_{\infty}$, the first contribution is already larger
than one. In other words, the infectiveness is so high that when
symptoms appear --and reconnection is turned on-- the infective
agent has already caused enough secondary infections to trigger the
spreading of the infection.

\begin{figure} %[!h]
\centerline{\includegraphics[width=12cm,clip=true]{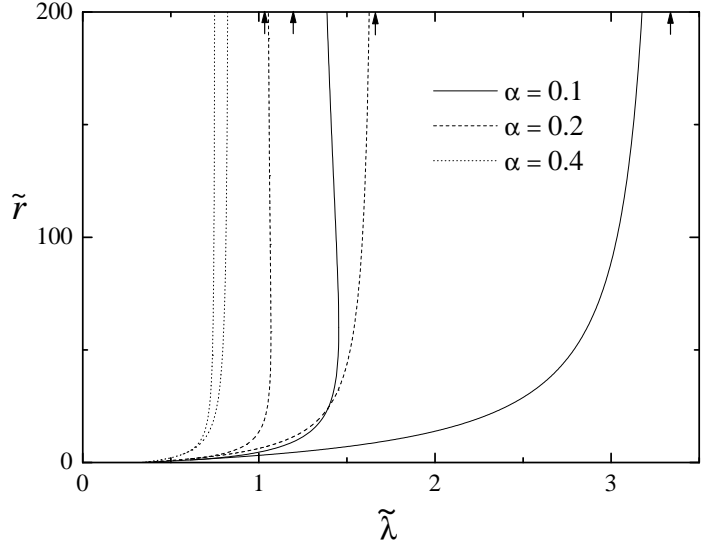}}
\caption{Phase diagram for a disease with an asymptomatic stage of
duration $\alpha / \gamma$, and three values of $\alpha$. For each
of them, the left-most and right-most curves represent  the tangent
and transcritical bifurcations, respectively. Compare with Figs.
\ref{figure2} and \ref{figure5}. The arrows indicate the position of
the vertical asymptotes of the curves for $\alpha=0.2$ and
$\alpha=0.4$: from right to left, they correspond to $\tilde
\lambda_1(\alpha=0.2)$, $\tilde \lambda_1(\alpha=0.4)$, $\tilde
\lambda_{\infty}(\alpha=0.2)$, $\tilde
\lambda_{\infty}(\alpha=0.4)$. No arrows have been drawn for
$\alpha=0.1$, because they would fall on top of the corresponding
curves.} \label{figure6}
\end{figure}

A numerical analysis of the equilibrium solutions of the system in
Eq. (\ref{ecs3}) reveals that, as before, a tangent bifurcation
occurs at a critical infectivity smaller that $\tilde \lambda_T$,
for sufficiently large reconnection rates. Figure \ref{figure6}
shows the phase diagrams for three values of $\alpha$. We see that
the critical infectivity corresponding to the tangent bifurcation
varies non-monotonically with $\tilde r$: it reaches a maximum
$\tilde \lambda_2$ at an intermediate value of the reconnection
rate, and then approaches a limit $\tilde \lambda_1$ as $\tilde r
\to \infty$, with $\tilde \lambda_1<\tilde\lambda_2 <\tilde
\lambda_T$.

Thus, the picture of the effects of reconnection is as follows. For
large values of infectivity, ${\tilde \lambda}> {\tilde
\lambda}_{\infty}$, the endemic state is stable for all ${\tilde
r}$. When the infectivity is in the interval $({\tilde \lambda}_2,
{\tilde \lambda}_T)$, increasing ${\tilde r}$ takes the system from
the endemic state to the bistable region, which here extends to
arbitrarily large values of the reconnection rate. Within this
region, as discussed in Section \ref{basic}, there is a threshold
initial density of infectives necessary to sustain the epidemics.
This breakpoint density grows with the reconnection rate and, for
$\tilde r \to \infty$, approaches a value smaller than the
corresponding equilibrium fraction of infectives.  Under these
conditions, even an arbitrarily large reconnection rate is unable to
suppress the infection, and high values of the initial fraction of
infectives asymptotically lead to the endemic state. For
infectivities in $({\tilde \lambda}_1, {\tilde \lambda}_2)$, there
is an intermediate interval of reconnection rates where the system
reaches the infection-free state. For sufficiently large values of
${\tilde r}$, though, the system returns to the bistable region.
Finally, for infectivities smaller than $\lambda_1$, the situation
is the same as in the previously discussed models: increasing
reconnection leads the system from the endemic state to the bistable
region and, upon further increase of $\tilde r$, the system reaches
the infection-free zone.

\begin{figure} %[!h]
\centerline{\includegraphics[width=12cm,clip=true]{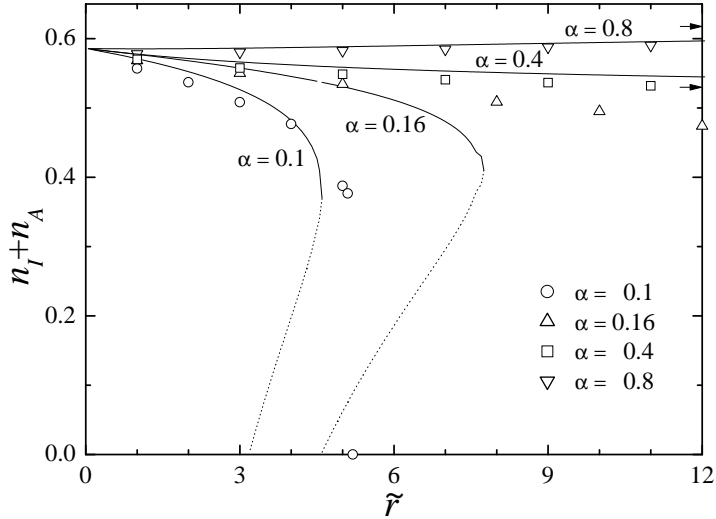}}
\caption{Bifurcation diagram for the total fraction of agents
carrying the infection, $n_I+ n_A$, for a disease with an
asymptomatic stage of duration $\alpha / \gamma$, for $k=3$, $\tilde \lambda=1$ and several values
of $\alpha$. Full and dashed lines represent stable and unstable equilibrium
solutions, respectively. Dots are simulation results for a system of
$5000$ agents with $\gamma=1$. The arrows indicate the asymptotic
value of $n_I$ for $\alpha=0.4$ and $\alpha=0.8$.} \label{figure7}
\end{figure}

\begin{figure} %[!h]
\centerline{\includegraphics[width=12cm,clip=true]{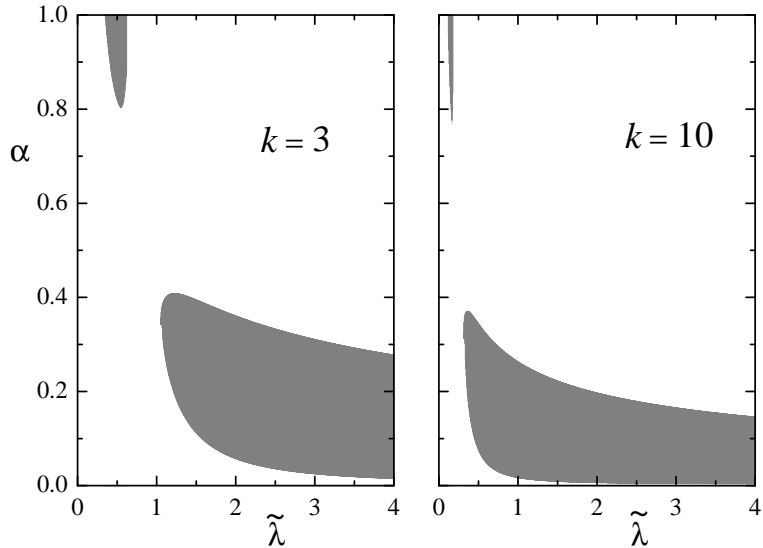}}
\caption{Phase diagrams showing, for two connectivities, the
regions (shaded) in parameter space ($\tilde \lambda$, $\alpha$)  where the
derivative of the fraction of agents that have contracted the
disease with respect to ${\tilde r}$, at ${\tilde r}=0$, is positive.} \label{figure8}
\end{figure}

Coming back to the question raised at the beginning of this section,
about the possibility of reconnection having unwanted effects in the
spread of the disease, the results shown in Fig. \ref{figure7}
suggest that the answer is not straightforward. As a function of the
reconnection rate $\tilde r$, the stable equilibrium fraction of
infectives can grow or decrease and, even for a given value of
$\tilde r$ and $\tilde \lambda$, the sign of the derivative of $n_I$
with respect to $\tilde r$ depends on $\alpha$. As a partial
characterization of this diversity of behavior, we analyze the sign
of $dn_I/d\tilde r$ for $\tilde r \to 0$, i.e. the slope of the
stable branches depicted in Fig. \ref{figure7} for small values of
the reconnection rate. Figure \ref{figure8} shows the zones of the
parameter plane $(\tilde \lambda, \alpha)$ where the equilibrium
fraction of infectives {\it increases} with the reconnection rate,
for $\tilde r \to 0$, and for two values of $k$. It can be shown
that, nevertheless, the growth of $n_I$ within this regions is small
if the infection level is itself low. In any case, in most of the
parameter plane, the fraction of infectives exhibit the desirable
decline of the infection level with reconnection.

It is tempting to apply these results to the propagation of real
diseases. The parameter $\alpha$ can be associated with  the number
of infections that occur before the onset of clinical symptoms
\citep{FRAF}. Using estimates of this number, we can have an idea of
how contact switching might work for these infections. For
gonorrhea, a rough estimate, using data from \citep{KvDS}, gives
$\alpha\approx 0.2$ and $\tilde \lambda \approx 4$. Being a sexually
transmitted disease, $k=3$ is a reasonable assumption. Using these
values, we see from the upper panel of Fig. \ref{figure8} that
gonorrhea would be in the zone where contact switching is not
effective. For influenza, the estimate gives $\alpha \approx 0.4$
\citep{FRAF}. Since, in contrast with gonorrhea, influenza is an
airborne disease, the number of contacts that may be infected is
large. Using $k=10$, the lower panel of Fig. \ref{figure8} shows
that, almost for any infectivity, contact switching might be an
effective strategy to curb the epidemic outbreak.  Needless to say,
any  application of these ideas should be based both on more
realistic models, taking into account specific features of the
disease under study, as well as on more sound estimates of the
parameters involved.

\section{Discussion and conclusion}
\label{conclusions}

We have studied the effects of contact switching as a strategy to
prevent the spread of an epidemic disease. Our analysis confirms,
and extends in several directions, the results obtained in previous
works \citep{GDdLB,ZR-G}. In particular, we have here relaxed some
important constraints, for instance, the assumption that every agent
knows the state of every other agent \citep{GDdLB}, and the
assumption that the mean number of contacts per agent is large
\citep{ZR-G}. In the basic model that we analyzed first, susceptible
agents with infective neighbors are allowed to reconnect these links
at rate $r$. For small values of this reconnection parameter, there
is a continuous transition --through a transcritical bifurcation--
from an infection-free state to an endemic state, as the infectivity
grows. When $r$ is larger than a certain critical value, on the
other hand, the transition becomes discontinuous and gives rise to a
bistability region where the infection can become asymptotically
extinct or endemic. In this situation, the asymptotic state is
determined by the initial fraction of infectives. Further increase
of the infectivity, however, makes the endemic state globally
stable, and the infection-free state becomes unstable.

One crucial feature of our basic model is that only susceptibles are
allowed to reconnect. Even though this constraint might seem
reasonable, the voluntary nature of contact switching suggests
relaxing it. We have analyzed a model where both types of agents can
switch contacts, with a new parameter $p$ giving the probability
that it is the infective agent who breaks and reconnects the link
with a susceptible neighbor. Remarkably, we find that, for certain
values of the parameters that characterize the disease, reconnection
can completely suppress the infection for intermediate values of
$p$, but not for larger or smaller values. Furthermore, even when
the disease is not completely suppressed, equilibrium infection
levels can be lower when more infectives are allowed to reconnect.
Even though, at first sight, this might seem counterintuitive, it
can be readily understood, at least for high infection levels: if
the fraction of infectives is large, susceptibles are likely to
reconnect their broken links to other infectives, thus  not altering
their risk of getting infected. On the other hand, when an infective
agent reconnects, which with high probability occurs to other
infectives,  the number of ``dangerous'' links decreases.

It has been shown that the outcome of many strategies to control
epidemic outbreaks depends not only on the reproductive number of
the disease, but also on the number of infections caused by an agent
before the onset of clinical symptoms \citep{FRAF}. In the case of
contact switching, it is even possible that the presence of an
asymptomatic phase leads to an acceleration of the spread of the
disease. In Section \ref{asymptomatics} we analyzed an extension of
the basic model, introducing a parameter $\alpha$ which gives, on
average, the fraction of time that the agent spends in an
asymptomatic phase during the infection. In this phase, the agent
behaves as a susceptible individual, but can transmit the infection
through contagion. We find that, in this case, reconnection does
lead to higher levels of infection, but this effect is restricted to
some limited regions of parameter space. Furthermore, within these
regions, the increase of this infection level due to reconnection is
rather small, unless its value is already very high in the absence
of reconnection.

It should be clear that any significant assessment of the benefits
and risks of contact switching must use sound estimates of all the
variables involved, applying them to realistic models of the
analyzed population. Important features that may be added to the
above models are, for instance, the population division into females
and males --who often exhibit rather different epidemiological
responses-- and age groups, as well as more complex degree
distributions in the initial network. Such generalizations, however,
will  hardly be amenable to analytical treatment. In our basic model
and its extensions, on the other hand, most epidemiological features
can be studied analytically. It is only after these have been fully
understood that more complex models should be tackled.

\appendix

\section{Appendix}
\label{appendix}

Models of epidemic spreading can be studied as stochastic contact
processes on  networks \citep{LD}. We illustrate the procedure with
the basic SIS model presented in Section \ref{basic}. For each site
$x$ of the network we introduce the probability  $P_t (A_x)$ that
the agent at $x$ is in state $A$ (either $S$ or $I$) at time $t$.
The evolution equation for this probability is:
\begin{equation}
\dot P_t(I_x)=- \dot P_t (S_x) =-\gamma P_t(I_x) + \lambda \sum_{y
\neq x} P_t(S_x \sim I_y) \label{eqP1}
\end{equation}
where $\gamma$ is the rate at which an infective agent becomes
susceptible, and $\lambda$ is the rate at which infective agents
infect each of their susceptible neighbors. Here, $P_t(S_x \sim
I_y)$ denotes the probability that the agents in sites $x$ and $y$
are susceptible and infective, respectively, and that they are
connected. Generally, we also introduce $P_t(A_x \sim B_y)$ as the
probability that an agent in state $A$ at site $x$ and an agent in
state $B$ at site $y$ are connected, and $P_t(A_x \nsim B_y)$ as the
probability that the same agents are {\sl not} connected. The
evolution equations for these two-sites probabilities are
\begin{eqnarray}
\dot P_t(S_x \sim I_y) &=& \gamma P_t(I_x \sim I_y) - \gamma  P_t(S_x \sim I_y) + \lambda \sum_{z \neq x,y}  P_t(S_x \sim S_y \sim I_z) - \nonumber \\ & & \lambda \sum_{z \neq x,y}  P_t(I_y \sim S_x \sim I_z) - \lambda  P_t(S_x \sim I_y) + \nonumber \\ & & r \sum_{z \neq x,y} \frac{P_t(I_y \nsim S_x \sim I_z)}{\sum_{v \neq z} P_t(S_y \nsim O_v)} - r \sum_{z \neq x,y} \frac{P_t(I_y \sim S_x \nsim O_z)}{\sum_{v \neq x} P_t(S_y \nsim O_z)}  \nonumber \\
\dot P_t(S_x \sim S_y) &=& \gamma ( P_t(I_x \sim S_y) + P_t(S_x \sim I_y)) - \nonumber \\ & & \lambda ( \sum_{z \neq x,y}  P_t(S_x \sim S_y \sim I_z) + \sum_{z \neq x,y}  P_t(S_y \sim S_x \sim I_z)  + \nonumber \\ & & r \left( \sum_{z \neq x,y} \frac{P_t(S_x \nsim S_y \sim I_z)}{\sum_{v \neq z} P_t(S_y \nsim O_v)} + \sum_{z \neq x,y} \frac{P_t(S_y \nsim S_x \sim O_z)}{\sum_{v \neq x} P_t(S_y \nsim O_z)} \right) \nonumber \\
\dot P_t (I_x \sim I_y) &=& -2 \gamma P_t(I_x \sim I_y) + \lambda (P_t(S_x \sim I_y)+(S_y \sim I_x) ) + \nonumber \\ & & \lambda \sum_{z \neq x,y } (P_t(I_x \sim S_y \sim I_z)+P_t(I_y \sim S_x \sim I_z)) \label{eq.system} \\
\dot P_t (S_x \nsim I_y) &=& \gamma P_t(I_x \nsim I_y) - \gamma  P_t(S_x \nsim I_y) +  \nonumber \\ & & \lambda \sum_{z \neq x,y}  P_t(S_x \nsim S_y \sim I_z) - \lambda \sum_{z \neq x,y}  P_t(I_y \nsim S_x \sim I_z) - \nonumber \\ & & r \sum_{z \neq x,y} \frac{P_t(I_y \sim S_x \nsim I_z)}{\sum_{v \neq z} P_t(S_y \nsim O_v)} - r \sum_{z \neq x,y} \frac{P_t(I_y \nsim S_x \sim O_z)}{\sum_{v \neq x} P_t(S_y \nsim O_z)} \nonumber \\
\dot P_t (I_x \nsim I_y) &=& -2 \gamma P_t(I_x \nsim I_y) +
\nonumber \\ & & \lambda \sum_{z \neq x,y } (P_t(I_x \nsim S_y \sim
I_z)+P_t(I_y \nsim S_x \sim I_z)) \nonumber
\end{eqnarray}
where $r$ is the rate at which a susceptible-infective link is
broken, and the susceptible agent reconnected to a random agent. The
remaining two-site probability  $P_t(S_x \nsim S_y)$ is obtained
from the fact that the sum of all two-sites probabilities equals
unity. The interpretation of the different summands in each equation
is rather straightforward. In the first equation,  for instance, the
first two terms account for the probability that a connected
susceptible-infective pair is created from an infective-infective
pair, or destroyed, by the spontaneous cure of an infective agent.
The third and fourth terms account for the probability that
infection is acquired from, or transmitted to, a neighbor outside
the  considered pair. The fifth term accounts for the probability
that infection is transmitted from the infective agent to the
susceptible agent in the same pair. The last two terms account for
the probability that the pair is created or destroyed by the
reconnection process. In these terms, the denominators take into
account the fact that, when the susceptible agent reconnects a link,
the new contact must exclude the previous neighbors of the agent in
question.

Within the {\em pair approximation} \citep{LD}, the three-site
probabilities in Eqs. (\ref{eq.system}) can be approximately given
in terms of two-site probabilities, thus closing the equation
system. The basic idea of this approximation is that, in a chain of
three of connected sites, $x \sim y \sim z$, the mutual interactions
between the central site $y$ of and each of the other two neighbors,
$x$ and $z$, can be decoupled:
\begin{equation}
P_t(A_x \sim B_y \sim C_z) \approx K \frac{P_t(A_x \sim B_y) P_t(B_y
\sim C_z)}{P_t(B_y)}.
\end{equation}
The constant $K$ is determined either from analytical
considerations, which give $K=(k-1)/k$, or from heuristic arguments
--for example, a better fit of simulations, as done here in the main
text.

We assume that all the probabilities are homogeneous over the
population, i. e. that they do not depend on the site, which is true
if the initial probabilities are also homogeneous. We also assume
that the epidemiological states at sites that are not connected can
be taken as independent, i. e. $P_t(A \nsim B) \approx P_t(A)
P_t(B)$ and $P_t(A \nsim B \sim B) \approx P_t(A) P_t(B \sim C)$
(using also the pair approximation). A further approximation that
simplifies the equations considerably is to assume that the number
of neighbors of each site is much smaller than the total number of
agents so that, when an agents reconnects, the probability that
another specific agent becomes the new contact is simply
$(N-1)^{-1}$, thus simplifying the denominator of the reconnection
terms.

To write the equations in terms of the fractions of agents, $n_S$ and
$n_I$, and links, $m_{SI}$, $m_{SS}$ and $m_{II}$ we take into
account  the  relations
\begin{eqnarray}
n_S(t) &=& P_t(S) ,\nonumber \\
m_{SS} (t) &=& \frac{N-1}{k} P_t(S \sim S),  \\
m_{SI} (t) &=& \frac{2(N-1)}{k} P_t(S\sim I),\nonumber
\end{eqnarray}
and $n_I=1-n_S$, $m_{II}=1-m_{SS}-m_{II}$. Using these equivalences,
and the approximations mentioned above, Eqs. (\ref{ecs1}) are
obtained from  system (\ref{eq.system}).

\end{document}